\documentstyle[12pt]{article}

\advance \topmargin by -\headheight
\advance \topmargin by -\headsep

\evensidemargin \oddsidemargin
\begin{document}

\newpage
\pagestyle{empty}
\centerline{\large\bf Lorentz Anomaly and 1+1-Dimensional Radiating
Black Holes\footnote{Work
supported in part by funds provided by
the U.S. Department of Energy (D.O.E.)
under cooperative agreement \#DE-FC02-94ER40818
and by the Istituto Nazionale di
Fisica Nucleare (INFN, Frascati, Italy).}}
\vskip 2 cm
\begin{center}
{\bf G. Amelino-Camelia$^{(a)}$, L. Griguolo$^{(b)}$,
and D. Seminara$^{(b)}$,}\\
\end{center}
\begin{center}
{\it (a)  Theoretical Physics, University of Oxford,
1 Keble Rd., Oxford OX1 3NP, UK} \\
{\it (b) Center for Theoretical Physics,
Laboratory for Nuclear Science, and Department of Physics,
Massachusetts Institute of Technology,
Cambridge, Massachusetts 02139, USA}
\end{center}

\vskip 4 cm
\centerline{\bf ABSTRACT }
The radiation from the black holes of a 1+1-dimensional
chiral quantum gravity model is studied.
Most notably,
a non-trivial dependence on a renormalization parameter
that characterizes the anomaly relations is uncovered
in an improved semiclassical approximation scheme;
this dependence is not present in the naive
semiclassical approximation.

\vfill
\noindent{OUTP-95-42-P \space\space\space
MIT-CTP-2484
\hfill October 1995}

\newpage
\pagenumbering{arabic}
\setcounter{page}{1}
\pagestyle{plain}
\baselineskip 12pt plus 0.2pt minus 0.2pt

The realization that the study of
some 1+1 dimensional quantum gravity
theories, like the CGHS model\cite{cghs},
might give the possibility to explore black hole quantum mechanics
in a rather simple (compared to the physical 3+1 dimensional context)
setting,
has led to great excitement.
(See Ref.\cite{bhlect}
for a review of related results.)
In particular,
there have been several attempts\cite{cghsmod}
to modify
the CGHS model in a way that
would allow a consistent analysis (from beginning to end)
of evaporating black holes.
These attempts have been only partly successful\cite{cghsmod},
but there might still be room for improvement
since just a limited class of modifications have been
considered.
Other recent papers have been devoted to
the investigation of the role of anomalies in the quantum
mechanics of 1+1 dimensional black holes.
Most notably, again in studies of the
CGHS model, it has been found\cite{navarri,as} that
the semiclassical
(quantized matter fields in a background geometry)
evaluation of the
Hawking radiation\cite{haw75} is essentially insensitive to
the addition of the local counterterm that converts the
Weyl anomaly into an anomaly
for diffeomorphisms of nonconstant Jacobian\cite{as,jack,abs95}.

In this Letter, we consider a modification of the CGHS model in which
chiral fermion fields replace the usual boson fields
as the matter content, and (accordingly) the geometry is described by the
{\it zweibein} $e_\mu^a$ instead of the metric $g_{\mu \nu}$.
($g_{\mu \nu} \! \equiv \! e_\mu^a e_\nu^b \eta_{ab}$
and $\eta \! \equiv \! diag(-1,1)$.)
We study the radiation of the black holes of this model
in the semiclassical approximation,
with particular attention to the role played by the
anomalies\footnote{We refer the reader to Ref.\cite{strochir}
for a discussion of certain other aspects
(different from the ones we are here concerned with)
of the physics of black holes in presence of chiral matter;
note, however, that in Ref.\cite{strochir} at the quantum level
only the special case with
equal number of left-moving fields and right-moving fields was
considered and therefore the structure of the anomalies was
much simpler.}.

Chiral quantum gravity theories
in 1+1-dimensions have a
rich anomaly structure, and no
inconsistency has been encountered
in their previous studies (see,
for example, Ref.\cite{chirqga,chirqgb});
it is therefore
quite natural to use one such theory for the investigation
of the role of anomalies in the quantum
mechanics of 1+1-dimensional black holes.
We also hope that the analysis here presented will motivate
future studies in which the features of
some chiral quantum gravity theories
be exploited in obtaining
a fully consistent description of black hole evaporation.

The model we consider has action $S_D+S^+_m+S^-_m$, where
\begin{eqnarray}
&&S_D = \int d^2 x \, E \,e^{-2 \phi} \left[
  R + 4 (\nabla \phi)^2 + 4 \lambda^2 \right]
{}~,
\label{sdilaton}\\
&&S^{\pm}_m =
- {1 \over 2} \sum_{n=1}^{N_{\pm}} \int d^2 x ~E~ e_a^\mu
\left (\bar\psi^{(\pm)}_n \gamma^a \frac{1\pm i\gamma_5}{2}
\partial_\mu\psi^{(\pm)}_n-
\partial_\mu\bar\psi^{(\pm)}_n\gamma^a \frac{1\pm i\gamma_5}{2}
\psi^{(\pm)}_n \right ).
\label{smatter}
\end{eqnarray}
$e_\mu^a$, $\phi$, and $\psi^{\pm}_n$ are
the {\it zweibein}, dilaton, and matter fields respectively.
$E$ denotes the determinant of the {\it zweibein}
(which equals $\sqrt{-det(g)}$).
The curvature $R$ can be written in terms of the spin-connection
$\omega_\mu = \epsilon^{\alpha \beta} e_\mu^a \eta_{ab}
\partial_\alpha e_\beta^b / E$ (we choose to work without
torsion),
as $R = 2 \epsilon^{\mu \nu} \partial_\mu \omega_\nu / E$.
Note that $S_D$,$S^{\pm}_m$ are invariant under
(tangent space) Lorentz
transformations and diffeomorphisms.
$S^\pm_m$ are also invariant under Weyl transformations.

With a straightforward generalization of the corresponding
analysis\cite{bhlect} of the CGHS model,
it is easy to verify that the classical equations of motion
of the model have black hole solutions.
In particular, the gravitational collapse of the left-moving shock wave
considered in Refs.\cite{cghs,bhlect,as} forms
the same classical black hole that it forms
in the CGHS model.
We shall
limit our analysis to this example of black hole,
so that our results can be more easily compared
to those of \cite{cghs,bhlect,as};
moreover, like in Refs.\cite{cghs,bhlect,as},
we introduce light-cone coordinates
and work in conformal gauge. [Note that,
since ours is a {\it zweibein theory},
the conformal gauge is characterized
by $e_\mu^\pm \! = \! e^\rho \delta_\mu^\pm$;
however, this
implies that $g_{+-} \! = \! - e^{2 \rho}/2$,
$g_{++} \! = \! g_{--} \! = \! 0$.]

For an
observer ${\hat e}_\mu^a(\sigma)$
that is {\it zweibein minkowskian}
(${\hat e}_\mu^a \! = \! \delta_\mu^a$)
at past null infinity\cite{as},
the solution of the classical equations of motion
that corresponds to the
black hole considered in Refs.\cite{cghs,bhlect,as}
has conformal factor\footnote{Like
in Refs.\cite{bhlect,as},
$a$ denotes the magnitude of the shock wave, and
the shock wave moves along $\sigma^+ \! = \! \sigma^+_0$.}
\begin{eqnarray}
{\hat \rho}  = - {1 \over 2}
\ln \left[1+ \Theta(\sigma^+ - \sigma^+_0)
{a \over \lambda} e^{\lambda \sigma^-}
\left( e^{\lambda (\sigma^+_0-\sigma^+)}-1 \right) \right]
{}~.
\label{background}
\end{eqnarray}

In the semiclassical analysis of Hawking radiation
one evaluates the quantum matter
energy-momentum tensor in the background geometry
of the black hole.
In our {\it zweibein} model the
matter energy-momentum tensor is
\begin{eqnarray}
{\hat T}^\mu_a \equiv - <\frac{1}{E} { \delta[S^+_m
+ S^-_m] \over \delta{e_\mu^a}} >_{\hat e} ~,
\label{tjack}
\end{eqnarray}
and for the example of black hole background that we are considering
the most interesting
physical information is encoded\cite{bhlect} in the value
of ${{\hat T}}_{--} \equiv {\hat T}^\mu_a {\hat e}_-^a {\hat g}_{\mu -}$
on\footnote{Like in Ref.\cite{bhlect,as},
${\cal I}_R^+$ (${\cal I}_R^-$) is
the future (past) null
infinity for right-moving light rays, and analogously
${\cal I}_L^+$ (${\cal I}_L^-$) is
the future (past) null
infinity for left-moving light rays.}
${\cal I}_R^+$.

\noindent
 From the results of Ref.\cite{chiranom} it follows
that the diffeomorphism
invariant quantization
of our chiral fields in a background geometry ${\hat e}_\mu^a$
leads to the anomaly relations (N.B.: ${\hat X} \equiv X({\hat e})$)
\begin{eqnarray}
{\hat \nabla}_\mu~  {\hat T}^\mu_a \!&=&\!-\frac{{\hat e}^\mu_a
{\hat \omega}_\mu}{24\pi}
\left (\frac{cq}{2} {\hat R}-2
h {\hat \nabla}_\mu {\hat \omega}^\mu\right )
{}~, \label{anod}\\
{\hat e}^a_\mu~ {\hat T}^\mu_a \!&=&\! \frac{1}{24\pi} \left [(c
+h) {\hat R}+ c q
{\hat \nabla}_\mu {\hat \omega}^\mu \right ]
{}~, \label{anow}\\
\epsilon^{a}_{~b} ~{\hat e}^b_\mu ~ {\hat T}^\mu_a  \!&=&\!
\frac{1}{24\pi} \left (\frac{cq}{2} {\hat R}-2 h
{\hat \nabla}_\mu {\hat \omega}^\mu\right )
{}~, \label{anol}
\end{eqnarray}
where $c \! \equiv \! (N_+ + N_-)/2$,
$q \! \equiv \! (N_+ - N_-)/(N_+ + N_-)$,
and $h$ is a free\footnote{Obviously, here ``free" should be
understood in the usual field theoretical sense, {\it i.e.} one
can choose freely among the values of $h$ consistent with unitarity
(for example in ordinary theories one cannot choose a negative mass
parameter). In Ref.\cite{chirqga} it was argued that only $h \! > \! 0$
would lead to a unitary chiral quantum gravity
theory; we shall come back to this later.}
renormalization parameter (the coefficient of a local
counterterm).
The emergence of the free parameter $h$ in chiral
quantum gravity theories
should be understood
in complete analogy with the
emergence of the Jackiw-Rajaraman parameter
in the quantization of the matter fields
of the Chiral Schwinger model\cite{chirschw}.

For arbitrary $c$,$q$,$h$ the right-hand
sides of Eqs.(\ref{anod}), (\ref{anow}), (\ref{anol})
do not transform covariantly under Lorentz and Weyl transformations,
also ${\hat T}^\mu_a$ does not transform covariantly
under these transformations. Instead, in spite of the unusual aspect
of Eq.(\ref{anod}) (that is due to the Lorentz anomaly),
${\hat T}^\mu_a$ transforms covariantly
under diffeomorphisms.
For $h \! = \! q \! = \! 0$ (diffeomorphism
and Lorentz invariant Dirac fermions)
the anomaly relations (\ref{anod}), (\ref{anow}), (\ref{anol})
are equivalent to the ones
encountered\cite{cghs,bhlect} in the CGHS model, and ${\hat T}^\mu_a$
transforms covariantly under both Lorentz transformations
and diffeomorphisms.
Also special is the case $q \! = \! 0$ (Dirac fermions), $h \! = \! -c$
which leads to a ${\hat T}^\mu_a$ that is traceless and  covariant
under diffeomorphism transformations, but Lorentz anomalous.

The Eqs.(\ref{anod}), (\ref{anow}), (\ref{anol})
can be used to evaluate ${\hat T}_{--}$ on $I_R^+$.
Indeed, these anomaly relations completely determine,
modulo the imposition of physical boundary conditions
at past null infinity,
the four components of ${\hat T}^\mu_a$.
This is analogous to the ordinary CGHS case, the only difference
being that here we have one more
component of the energy-momentum tensor (the antisymmetric component
coming from the Lorentz anomaly)
and one more anomaly
relation (the one expressing the Lorentz anomaly).

Imposing as
physical boundary condition\cite{bhlect,as} that ${\hat T}^\mu_a$  vanish
on ${\cal I}_L^-$ ({\it i.e.} $\sigma^+ \! = \! - \infty$),
and that there be no incoming radiation
along ${\cal I}_R^-$ ({\it i.e.} $\sigma^- \! = \! - \infty$) except
for the classical shock-wave at $\sigma^+ \! = \! \sigma^+_0$,
with a straightforward (but lengthy) derivation
one finds that
\begin{eqnarray}
{\hat T}_{--} = {1 \over 12 \pi}
\left[ \left (c+h - \frac{c q}{2}\right )
\left( \partial^2_{\sigma^-} \hat\rho -
(\partial_{\sigma^-} \hat \rho)^2\right)
-\frac{cq}{2} (\partial_{\sigma^-} \hat \rho)^2\right]
{}~,
\label{solstroconfd}
\end{eqnarray}

The physical interpretation of the energy flux
seen on ${\cal I}_R^{+}$
is clearest\cite{bhlect} to observers that are asymptotically
minkowskian on ${\cal I}_R^{+}$, but
from Eq.(\ref{background}) one sees that the ${\hat e}_\mu^a(\sigma)$
observer is not minkowskian on ${\cal I}_R^{+}$ (whereas it is
minkowskian on ${\cal I}_{R,L}^{-}$).
An observer ${\check e}_\mu^a(y)$ that
is {\it zweibein minkowskian} on ${\cal I}_R^{+}$ can be
obtained from ${\hat e}_\mu^a(\sigma)$ by combining
the (conformal) diffeomorphism $\sigma^\pm \! \rightarrow \! y^\pm$,
where
\begin{eqnarray}
y^+ = \sigma^+ ~,~~~ y^- = - \ln(e^{- \lambda \sigma^-}
- a/\lambda)/\lambda
{}~,
\label{coordrede}
\end{eqnarray}
and the Lorentz
transformation ${\check e}_\mu^\pm(y) \! \rightarrow \! (1
+ a e^{\lambda y^-}/\lambda)^{\pm 1/2} \, {\check e}_\mu^\pm(y)$.

\noindent
Using the fact that the energy-momentum tensor
transforms covariantly under
diffeomorphisms, but under Lorentz
transformations it follows the
transformation rules implied by
the anomaly relations (\ref{anod}), (\ref{anow}), (\ref{anol}),
one finds that the observer ${\check e}_\mu^a(y)$
sees the following ${\check T}_{--}$
\begin{eqnarray}
{\check T}_{--}(y) = [{\hat T}_{--}(\sigma(y))
+ \Delta^{{\hat e}\rightarrow {\check e}}_{--}(\sigma(y))]
(d\sigma^- / dy^-)^2
\label{ttrasf}
\end{eqnarray}
where
\begin{eqnarray}
\Delta^{{\hat e}\rightarrow {\check e}}_{--} \! = \!
\frac{1}{48 \pi}
\left\{
h \, \left[ \partial_{\sigma^-}^2 \! \left(  \ln{dy^- \over
d\sigma^-} \right ) \right]^2      -      \left( 2 h + c q \right )
\partial_{\sigma^-}^2 \! \left(  \ln{dy^- \over d\sigma^-} \right )
+ 2 c q \,
\partial_{\sigma^-} \hat \rho \,\,
\partial_{\sigma^-} \! \left(  \ln{dy^- \over d\sigma^-} \right ) \right\}
{}~.
\label{deltamunu}
\end{eqnarray}
Since on ${\cal I}^+_R$
({\it i.e.} $y^+ \! = \! \infty$)
one finds that ${\hat \rho} \! = \! (1/2) \ln (dy^- / d \sigma^-)$,
from Eqs.(\ref{solstroconfd})-(\ref{deltamunu})
it follows that ${\check T}_{--}$ is $h$-independent on ${\cal I}^+_R$
\begin{eqnarray}
\left[{\check T}_{--}\right]_{{\cal I}^+_R} = c (1 - q)
{\lambda^2 \over 48 \pi}
\left[ 1 - (1+a e^{\lambda y^-}/\lambda)^{-2} \right]
{}~.
\label{confdeltaty}
\end{eqnarray}

Notice that
this result reproduces the corresponding result\cite{cghs,bhlect}
for the CGHS model
upon replacing $c (1 - q) \! = \!N_-$ with $N$, the number
of boson fields of the CGHS model.
This might indicate that, in spite of the Lorentz anomaly,
in the present approximation
left-movers and right-movers are still decoupled\cite{bhlect};
in fact, such a decoupling implies that
the Hawking radiation on ${\cal I}^+_R$ due to our black hole
(formed by gravitational collapse of a left-moving shock wave)
should only be sensitive to $N_-$.

The fact that the complicated structure of the anomalous
Lorentz transformations of the energy momentum tensor
and the anomaly relations
(\ref{anod}), (\ref{anow}), (\ref{anol}),
conspire to give the
$h$-independent result (\ref{confdeltaty})
is consistent with the findings of Ref.\cite{as}, where the
semiclassical analysis of the black hole radiation
in the ordinary CGHS model
was shown to be insensitive to the value of
the coefficient of a local regularization counterterm.
However, the similarities between our model
and the Chiral Schwinger model (ours is essentially a gravitational
version of the Chiral Schwinger model)
suggest that the parameter $h$ should have
some non-trivial physical role;
for example, the Jackiw-Rajaraman parameter
affects the value of the mass emergent\cite{chirschw} in the
Chiral Schwinger model.
Moreover,
investigations\cite{chirqgb}
of related chiral quantum gravity theories have found
that $h$ does have a non-trivial
role in the fully quantized theory;
most notably, the central charge has been found to depend on it.
We therefore expect that in the fully quantized theory
the black hole radiation
depends on $h$ even though this
is not seen in the naive semiclassical analysis.
To test this expectation, in the following
we shall go beyond the naive semiclassical
approximation; specifically, we shall exploit the fact that
it is rather simple to perform the functional integration
over one of the four fields characterizing the {\it zweibein},
and include the corresponding quantum correction in our analysis.

A general {\it zweibein} $e_\mu^a$ can be written in terms
of a field $\Upsilon$, which we shall
call {\it Lorentzon}, and a Lorentz
gauge-fixed {\it zweibein} $[{\hat e}_\mu^\pm]_{\hat \Upsilon}$
({\it i.e.} a {\it zweibein} whose
orientation ${\hat \Upsilon}$ in the
tangent space is prescribed, and therefore
is characterized by only three
fields),
as follows
\begin{eqnarray}
e_\mu^\pm = e^{\pm \Upsilon} [e_\mu^\pm]_{\hat \Upsilon}
{}~.
\label{lorentzon}
\end{eqnarray}
In our theory it is rather simple to integrate
out\footnote{The {\it Lorentzon} corresponds
to the degree of freedom ``turned on" by the anomaly.
The consequences of integrating it out are therefore particularly
interesting.}
the field $\Upsilon$;
in fact,
the corresponding functional integration can be cast in Gaussian
form\cite{chirqgb}, and
choosing a diffeomorphism and Lorentz invariant measure
of the type proposed in Ref.\cite{stroghost},
one finds that,
as we shall show in detail in Ref.\cite{agsprep},
the {\it Lorentzon}-integrated
energy-momentum tensor satisfies
the following anomaly relations\footnote{Note that these
anomaly relations
receive no contribution from the {\it Lorentzon} measure and
the Lorentz ghosts. If, instead of following the proposal of
Ref.\cite{stroghost},
one defines the measures in the ordinary
way (with the real metric), the Lorentz ghosts
still do not contribute to the anomaly relations, but
a contribution from
the {\it Lorentzon} measure appears.
Since this contribution can be reabsorbed
by increasing the value of $c$ by $1$,
our analysis would not be qualitatively modified.}
in the black hole background ${\hat e}^a_\mu$
\begin{eqnarray}
{\hat \nabla}_\mu  {\hat T}^\mu_a \!&=&\! 0 ~, \label{anodl}\\
{\hat e}^a_\mu~ {\hat T}^\mu_a  \!&=&\! [c + h +
c^2 q^2 /(4h)] {\hat R}/(24 \pi) ~, \label{anowl}\\
\epsilon^{a}_{~b} ~{\hat e}^b_\mu ~{\hat T}^\mu_a \!&=&\! 0
{}~. \label{anoll}
\end{eqnarray}
One can recognize that these anomaly relations become
equivalent to the ones
encountered\cite{cghs,bhlect} in the ordinary CGHS model, upon
replacing $c_{eff} \equiv c + h + c^2 q^2/(4h)$ with $N$.
Therefore, once the
{\it Lorentzon} is integrated out,
the semiclassical analysis of black hole radiation in our model
can be performed exactly as in the ordinary CGHS model
(in particular the {\it Lorentzon}-integrated ${\hat T}^\mu_a$
transforms covariantly under both Lorentz transformations
and diffeomorphisms); this leads to the result that for
the ``good observer" ${\check e}_\mu^a(y)$
\begin{eqnarray}
\left[{\check T}_{--}\right]_{{\cal I}^+_R} =
\left[ c + h + {c^2 q^2 \over 4h} \right]
{\lambda^2 \over 48 \pi} \left[ 1
- (1+a e^{\lambda y^-}/\lambda)^{-2} \right]
{}~.
\label{finalresult}
\end{eqnarray}
Hence, we find
again that the black hole radiation on ${\cal I}^+_R$
has the same functional dependence on $y^-$ as in the
semiclassical analysis\cite{cghs,bhlect} of the
CGHS model, but in the present case the overall coefficient
has an $h$-dependence.

The specific dependence on $h$ of
the overall coefficient $c_{eff}$
is quite interesting.
In particular, $c_{eff}$ diverges at $h\! = \! 0$, and
$c_{eff}\! \le \! N_-$
whenever $h \! < \! 0$;
this appears to be in agreement with the expectation that the
theory be not unitary\footnote{Work is in progress\cite{agsprep}
attempting to test more rigorously
whether the theory with negative $h$ is physical
at least when $c_{eff}\! \ge \! 0$, {\it i.e.}
$-c [1 + (1 - q^2)^{1/2}]/2 \le h \le -c [1 - (1 - q^2)^{1/2}]/2$.}
(the {\it Lorentzon}\footnote{Note that,
in the spirit of Ref.\cite{stroghost}, we have included
no contribution from the measures of the
{\it Lorentzon}, dilaton, conformal factor, and ghosts.
However, the (diffeomorphism invariant)
quantization of the chiral matter fields (whose measure has
to involve the physical metric\cite{stroghost})
still induces
a kinetic term for the {\it Lorentzon}.}
is ghost-like\cite{chirqga})
when $h \! \le \! 0$.
Instead,
$c_{eff}\! \ge \! N_-$
(consistently with the absence\cite{chirqga} of ghost-like fields)
whenever $h \! > \! 0$.

Our result that after integrating out the
{\it Lorentzon} the radiation has an $h$-dependence,
which is not present in
the naive semiclassical approximation,
also allows to
reconcile the findings of Refs.\cite{navarri,as}
with the ones of Refs.\cite{chirqga,chirqgb,chirschw}.
Notably,
the ${\check T}_{--}$ in Eq.(\ref{finalresult})
takes the same value in correspondence of pairs of values of $h$,
just like
paired values of the Jackiw-Rajaraman parameter
lead to the same value of the mass emergent in the Chiral Schwinger
model\cite{chirschw}.

In conclusion, we want to emphasize that in this paper we have
only explored some aspects of a torsionless
chiral quantum gravity theory
which are relevant to the issue of the possible role of
renormalization parameters (coefficients of local counterterms)
in 1+1-dimensional black hole radiation;
however, it is our expectation that other problems
of 1+1-dimensional black hole quantum mechanics might
be investigated by exploiting the rich structure of
chiral quantum gravity theories.
For example, it would be interesting to check whether
a fully consistent analysis of the back reaction\cite{cghsmod}
(a necessary step
toward a rigorous description of black hole evaporation,
which has been attempted without success in the CGHS model)
is possible in some (not necessarily torsionless)
chiral quantum gravity theory.

\bigskip
\bigskip
We happily acknowledge conversations with R. Jackiw, E. Keski-Vakkuri,
S. Mathur, M. Ortiz, and B. Zwiebach.

\newpage
\baselineskip 12pt plus .5pt minus .5pt

\end{document}